\documentclass{tcibook}
\usepackage{fancyhea}
\usepackage{work}
\usepackage{bm}       
\usepackage{graphicx}
\usepackage{multirow}
\usepackage{lineno}
\usepackage{threeparttable}
\usepackage[normalem]{ulem}
\usepackage{color}
\usepackage[colorlinks = true,
            linkcolor = blue,
            urlcolor  = blue,
            citecolor = blue,
            anchorcolor = blue]{hyperref}

\def\beq{\begin{equation}}
\def\eeq#1{\label{#1}\end{equation}}
\def\eeqn{\end{equation}}

\newenvironment{Eqnarray}%
   {\arraycolsep 0.14em\begin{eqnarray}}{\end{eqnarray}}
\def\beqa{\begin{Eqnarray}}
\def\eeqa#1{\label{#1}\end{Eqnarray}}
\def\eeqan{\end{Eqnarray}}

\let\bar=\overbar

\def\lsim{\mathrel{\raise.3ex\hbox{$<$\kern-.75em\lower1ex\hbox{$\sim$}}}}
\def\gsim{\mathrel{\raise.3ex\hbox{$>$\kern-.75em\lower1ex\hbox{$\sim$}}}}

\def\del{\partial}
\def\Dslash{\not{\hbox{\kern-4pt $D$}}}
\def\dslash{\not{\hbox{\kern-2pt $\del$}}}
\def\pslash{\not{\hbox{\kern-2pt $p$}}}
\def\ETmiss{\not{\hbox{\kern-4pt $E$}}_T}

\def\Dlr{\mathrel{\raise1.5ex\hbox{$\leftrightarrow$\kern-1em\lower1.5ex\hbox{$D$}}}}

\def\MSB{{\bar{M \kern -2pt S}}}
\def\msb{{\bar{\scriptsize M \kern -1pt S}}}

\def\drb{{\bar{\scriptsize D \kern -1pt R}}}

\def\authorlist#1#2{
    \vskip 0.4in
\begin{center}\begin{large} {\bf  #1 } \end{large}
    \vskip 0.2in
              #2
     \vskip 0.2in
   \end{center}
}

\begin{document}


\pagenumbering{roman}

\parindent=0pt
\parskip=8pt
\setlength{\evensidemargin}{0pt}
\setlength{\oddsidemargin}{0pt}
\setlength{\marginparsep}{0.0in}
\setlength{\marginparwidth}{0.0in}
\marginparpush=0pt


\pagenumbering{arabic}

\renewcommand{\chapname}{chap:intro_}
\renewcommand{\chapterdir}{.}
\renewcommand{\arraystretch}{1.25}
\addtolength{\arraycolsep}{-3pt}

































 \setcounter{chapter}{9} 


\chapter{Radio Detection}

\authorlist{A. Connolly, A. Karle}
   {Sijbrand de Jong, Cecil Thomas}


\section{IF 10: Executive Summary}

Detection techniques at radio wavelengths
play an important role in the future of 
astrophysics experiments.
The radio detection of cosmic rays, neutrinos, and photons has emerged as the technology of choice at the highest energies.   Cosmological surveys require the detection of radiation at mm wavelengths at thresholds down to the fundamental noise
limit.

High energy astroparticle and neutrino detectors use large volumes of a naturally occurring suitable dielectric: the Earth's atmosphere and large volumes of cold ice as available in polar regions.
The detection technology for radio detection
of cosmic particles has matured in the past decade and is ready to move beyond prototyping or midscale applications. 
Instrumentation for radio detection has reached a maturity for science scale detectors. 
Radio detection provides competitive results in
terms of the measurement of energy and direction and in particle identification when to compared to currently applied technologies for high-energy
neutrinos when deployed in ice and for
ultra-high-energy cosmic rays, neutrinos, and
photons when deployed in the atmosphere.  It 
has significant advantages in terms of cost per
detection station and ease of deployment.

\begin{itemize}
\item IF10-1  While conceptual designs exist for next-generation arrays, investment in R\&D can reduce cost and optimize designs.
\item IF10-2  Opportunities exist in optimizing for power and simplifying: e.g. by using ASIC-based digitizer/readout (PMT) or RFSoCs (Radio Frequency System on Chip).  Investigations are needed to identify synergies with experimental needs in other areas.
\item IF10-3  Remote power and communications approaches of very large extended arrays can still benefit from dedicated R\&D. Explore synergies with experiments in other areas, like DUNE, SKA, CTA, which also need large distance communication and synchronization.
\item IF10-4  Enable future mm-wave cosmic probes through R\&D and pathfinder experiments of new mm-wave detectors with higher channel density relative to current CMB detectors.
\end{itemize}

\section{Radio detection of cosmic particles}

Astrophysical neutrinos, cosmic rays, and gamma rays are excellent probes of astroparticle physics and high-energy physics
~\cite{NF-04-Ackerman}. High-energy and ultra- high-energy cosmic particles probe fundamental physics from the TeV scale to the EeV scale and beyond.

Radio detection offers opportunities to
instrument large areas on the surface of the earth.
This is important because at
energies above an EeV, the cross section for
neutrinos interacting with matter have increased
so much that the detector area becomes a more
important parameter for the acceptance then
the fiducial volume.  At these energies, the 
preferred conversion targets for the neutrinos
are mountains, the earth's crust, ice, or even the
atmosphere.  
For ultra-high energy cosmic rays and photons,
radio detection utilizes the air shower formed in the atmosphere.

While the optical Cherenkov technique has been enormously successful in
measuring neutrinos from $\mathcal{O}$({MeV}) to $\mathcal{O}$(10\,{PeV}), 
at higher energies, techniques at radio frequencies are the most promising for neutrino detection. 
In the energy range from 10 TeV to 30 PeV a factor of 5 in increase 
of sensitivity is needed in order to move from the discovery of cosmic 
neutrinos and the first correlation with sources to 
identify the sources of the cosmic neutrino flux and explore fundamental physics 
with high energy neutrinos. 

Like in the optical, for radio techniques the two media being pursued broadly are the atmosphere and solid natural ice, the latter with emphasis on neutrinos. Both can be used to pursue science with cosmic rays, energetic neutrinos, or both.  
Above about 100\,PeV, volumes of order 1000\,km$^{3}$, which can only be naturally occurring, are
necessary to detect the rare astrophysical neutrino flux in this regime.  Volumes of this scale are too large to instrument with the tens-of-km spacing necessary for optical sensors. 
This spacing of optical is set by the distance over which optical light is absorbed or scattered in the glacial ice that is used as the detection medium.  
However, clear ice is transparent
to radio-frequency signals over distances
of order 1\,km, which sets the 
typical spacing for  detectors
using radio techniques.

There are many
different approaches to the detection
of cosmic particles at radio frequencies~\cite{NF-04-Wissel}.  These approaches, the instrumentation that is unique and common to each, and
future developments are described in this section.

\subsection{Askaryan}

In matter, particle cascades induced by cosmic particles produce Askaryan emission. The emission comes about
from the time evolution of a charge
asymmetry that develops in these particle 
cascades by ionization electrons moving with the
cascade's front and positive ions staying behind.  
It is coherent
for frequencies up to about 1\,GHz when viewed
at the Cherenkov angle
and is linearly polarized perpendicular to
the cascade direction and inward to its axis.
Askaryan emission was measured in test beam
experiments in the 2000s.  It has also been observed
to be emitted from cosmic ray air showers.
 Now, many
experiments are
searching for this signature from 
ultra-high energy neutrino interactions 
either from within the ice or by viewing ice from altitude.

\subsubsection{In-ice}

One approach to detecting neutrinos via
the Askaryan signature is by embedding
antennas into the ice itself, and the
first experiment to take this approach
was RICE.  Since then, ARIANNA, ARA, and
RNO-G~\cite{NF-04-RNOG-Wissel} search for ultra-high energy 
neutrinos with antennas deployed
in the ice, either at shallow depths (within $\sim$10\,m of the surface) or somewhat deeper (up to 200\,m).  An integrated facility for 
a wide band neutrino detector is IceCube-Gen2 that will employ both optical and radio detectors.  Deep detectors are
employing interferometry-based trigger designs.


\subsubsection{Balloon}

Another approach to detecting the
Askaryan signature from ultra-high energy 
neutrinos is to deploy antennas on 
a balloon-borne
payload at stratospheric altitudes where
the payload can view 
approximately 1.5\,million km$^2$ of ice.
At these altitudes, the threshold for a detection
is higher, but above threshold, balloons greatly exceed
ground-based approached in viewable area.
The ANITA project flew four times
under NASA's long-duration balloon
program, and the next-generation PUEO~\cite{NF-04-Vieregg}
is set to launch in the 2024-2025 Austral 
summer season with a trigger that used an interferometric
approach.

\section{Geomagnetic emission}

A cosmic ray incident on the atmosphere will
produce a particle cascade, and charges in
the cascade produce a transverse current due to the earth's magnetic field, which produces 
what is known as geomagnetic emission at radio
frequencies.  The geomagnetic emission is coherent
up to frequencies of about 1\,GHz
and is linearly polarized perpendicular to both the
earth's magnetic field and cascade direction.
The Askaryan emission mentioned above also
contributes to the radio frequency signal from
air showers on average at the 20-30\,\% level in energy
density.  The interplay between geomagnetic
and Askaryan emission gives rise to a non-azimuthally symmetric signal pattern due to
the interplay of the different radiation patterns.
Both are collimated in a Cherenkov cone with
an opening angle of typically a degree.

A next generation ultra-high energy cosmic
particle observatory is required to exploit 
proton astronomy and to discover ultra-high
energy neutrinos and photons.  Such a next-generation cosmic particle observatory
needs to cover a large area, up to 200,000\,km$^2$.
Radio detection is the current favorite to make
deployment at this scale possible and GRAND
proposed to cover an area of this size.

\subsection{Radio detection of cosmic rays}

In addition to the traditional strategies
for detecting high energy cosmic rays incident on
the atmosphere, which include detection
of secondary particles, optical Cherenkov
emission, and fluorescence emission,
geomagnetic emission is a strategy that
complements the others and
has seen major advancements in the past two decades
~\cite{CF-07-Schroeder}.  Radio detection of cosmic rays
has a round-the-clock duty cycle and leaves
a broad ($\sim 100$\,m) footprint on the ground.

\subsection{Detection of neutrino-induced, earth-skimming air showers}

The combined geomagnetic and Askaryan signatures can also be used to 
detect air showers that can originate from
energetic tau neutrinos that are earth skimming~\cite{CF-07-Abraham}.  These can produce a tau
lepton through a charged current interaction
in matter, and the tau can subsequently decay to produce an electromagnetic or hadronic shower.

The detection of pulses consistent with air showers going in the upward
direction  has the advantage of being flavor sensitive, and has become an objective of many 
detectors.
BEACON~\cite{CR-07-Wissel}, in prototype phase, TAROGE, and TAROGE-M are compact antenna arrays in elevated locations that aim to detect UHE $\nu_{\tau}$ emerging upwards via the radio emission of the air showers that they trigger. ANITA  and PUEO  are also sensitive to upgoing $\nu_{\tau}$ , from a higher elevation. GRAND~\cite{NF-04-Bustamante} is a planned experiment that will cover large areas with a sparse antenna array to detect the radio emission from air showers triggered by UHE 
$\nu_{\tau}$, cosmic rays, and gamma rays.
This is also a neutrino signature for PUEO, the
follow-up of ANITA, and many other dedicated projects instrumenting large areas with
antennas such as GRAND or putting them in mountains such as BEACON
and TAROGE.

\subsection{RADAR}

An alternate strategy for detection
of neutrinos at radio frequencies uses RADAR~\cite{IF10-Prohira}.
When a high-energy neutrino interacts in the ice, it produces a relativistic cascade of charged particles that traverse the medium. As they progress, they ionize the medium, leaving behind a cloud of stationary charge. This cloud of charge, which persists for a few to tens of nanoseconds, is dense enough to reflect radio waves. Therefore, to detect a neutrino, a transmitter can illuminate a volume of dense material like ice, and if a neutrino interacts within this volume, the transmitted radio will be reflected from the ionization cloud to a distant receiver, which monitors the same illuminated volume.

With this technique, a custom signal is transmitted in the ice and received 
after reflections from neutrino-induced
cascades.  With this technique, the
experimenter can determine the properties
of the signal (including the amplitude,
up to what is permitted to be transmitted).
Also, the radar method has excellent geometric acceptance relative to passive 
(Askaryan) methods, which require the detector to lie within a small angular window 
at the Cherenkov angle.  Recent test beam measurements have demonstrated the feasibility of the method in the laboratory, with in-situ tests forthcoming. RET-CR will serve as a pathfinder experiment, and RET-N could make radio detections of UHE neutrinos within the decade with the potential to complement or improve upon existing technologies in this
energy regime.


\section{KIDs}

Cosmological surveys require the detection of radiation at mm wavelengths at thresholds down to the fundamental noise
limit.  The detection of mm-wave radiation is
important for: studying cosmic acceleration (Dark Energy) and testing for deviations from general relativity expectations through measurements of the kinetic Sunyaev-Zeldovich effect, precision cosmology (sub-arcminute scales) and probing new physics through ultra-deep measurements of small-scale CMB anisotropy, and mm-wave spectroscopy to map out the distribution of cosmological structure at the largest scales and highest redshifts.

Imaging and polarimetry surveys at sub-arcminute scales will require O($10^6$) detectors over a O(10 deg$^2$) fields-of-view (FoV) covering 9 spectral bands from 30\,GHz to 420\,GHz. Spectroscopic surveys (over a smaller FoV initially, O(1\,deg$^2$), but potentially also reaching O(10\,deg$^2$)) will require a further factor of 10–100 increase in detector count.
The 2019 report of the DOE Basic Research Needs Study on High Energy Physics Detector Research and Development 
identified the need to carry out detector R\&D to 
achieve this goal.
The driver for new technology is not the detector count
but the increased detector {\it density}.   Whereas in
current experiments, the detector packing density is
limited by the physical size of elements in
existing demonstrated multiplexing schemes, KIDs
eliminate the need for additional cold multiplexing
components, allowing for arrays at the densities needed
for the science aims of proposed cosmological surveys.

The kinetic inductance detector (KID) is a technology that has gained significant traction in a wide range of applications across experimental astronomy over the last decade. A KID is a pair-breaking detector based on a superconducting thin-film microwave resonator, where the relative population of paired (Cooper pairs) and un-paired (quasiparticles) charge carriers govern the total complex conductivity of the superconductor. Photons with energy greater than the Cooper pair binding energy (2$\Delta$) are able to create quasiparticle excitations and modify the conductivity. By lithographically patterning the film into a microwave resonator, this modification is sensed by monitoring the resonant frequency and quality factor of the resonator.
Since each detector is formed from a microwave resonator with a unique resonant frequency,  a large number of detectors can be readout without the need for additional cryogenic multiplexing components.  In addition, the designs to be fabricated are 
relatively simple.

There are a number of KID-based architectures being developed  for a variety of scientific applications.  Direct Absorbing KIDs are
the simplest variant, where the resonator geometry is optimized to act as an impedance-matched absorber to efficiently collect the incoming signal.  To date, the only facility-grade KID-based instruments are based on this detector architecture, with the NIKA-2 experiment on the IRAM 30-meter telescope having demonstrated that the KID-based instruments are highly competitive with other approaches.  Microstrip-coupled KIDs take
advantage of recent advancements that
allow for
the ability to lithographically define 
circuits capable of on-chip signal processing with extremely low loss.  The capability
to robustly couple radiation from superconducting thin-film microstrip 
transmission lines into a KID with high optical efficiency is being developed.
Thermal Kinetic Inductance Detectors take
a similar approach as has been developed
for bolometric transition edge sensor arrays.
Instead of directly absorbed radiation breaking pairs, a thermally-mediated KID (TKID) uses the intrinsic temperature response of the superconducting film to monitor the temperature, and therefore absorbed power.
It combines the multiplexing advantage of
KIDs with the proven performance of bolometric designs in TES detectors, at the expense of fabrication complexity.

On-chip spectroscopy is a
natural extension of multi-band imaging using on-chip filters  to a filter-bank architecture to realise medium-resolution spectroscopic capability.  Several approaches to on-chip spectroscopy exist at a range of technological readiness.

\end{document}